\documentclass[10pt,prd,twocolumn,superscriptaddress,showpacs,amsmath,nofootinbib]{revtex4-1}

\usepackage[english]{babel}
\usepackage{graphicx}
\usepackage{amssymb}
\usepackage{mathtools}

\usepackage[unicode=true,breaklinks=true,colorlinks=true,allcolors=blue]{hyperref}
\hypersetup{pdftitle=Precise determination of the branching ratio of the neutral-pion Dalitz decay}
\hypersetup{pdfauthor=Tom\'{a}\v{s} Husek}

\makeatletter
\providecommand*{\diff}%
  {\@ifnextchar^{\DIfF}{\DIfF^{}}}
\def\DIfF^#1{%
  \mathop{\mathrm{\mathstrut d}}%
    \nolimits^{#1}\gobblespace}
\def\gobblespace{%
    \futurelet\diffarg\opspace}
\def\opspace{%
    \let\DiffSpace\!%
    \ifx\diffarg(%	
      \let\DiffSpace\relax
     \else
      \ifx\diffarg[%
	\let\DiffSpace\relax
      \else
	\ifx\diffarg\{%
	  \let\DiffSpace\relax
	\fi\fi\fi\DiffSpace}

\begin{document}

\title{Precise determination of the branching ratio of the neutral-pion Dalitz decay}
\author{Tom\'{a}\v{s} Husek}
\email{thusek@ific.uv.es}
\affiliation{
IFIC,
Universitat de Val\`encia -- CSIC,
Apt.\ Correus 22085, E-46071 Val\`encia, Spain}
\author{Evgueni Goudzovski}
\email{eg@hep.ph.bham.ac.uk}
\affiliation{
School of Physics and Astronomy,
University of Birmingham,
Edgbaston, Birmingham, B15 2TT, United Kingdom}

\author{Karol Kampf}
\email{karol.kampf@mff.cuni.cz}
\affiliation{
Institute of Particle and Nuclear Physics,
Faculty of Mathematics and Physics,
Charles University,
V Hole\v{s}ovi\v{c}k\'{a}ch 2, 18000 Praha 8, Czech Republic}

\begin{abstract}
We provide a new value for the ratio $R={\Gamma(\pi^0\to e^+e^-\gamma(\gamma))}/{\Gamma(\pi^0\to\gamma\gamma)}=11.978(6)\times10^{-3}$, which is by two orders of magnitude more precise than the current Particle Data Group average.
It is obtained using the complete set of the next-to-leading-order radiative corrections in the QED sector, and incorporates up-to-date values of the $\pi^0$-transition-form-factor slope.
The ratio $R$ translates into the branching ratios of the two main $\pi^0$ decay modes: $\mathcal{B}(\pi^0\to\gamma\gamma)=98.8131(6)\,\%$ and $\mathcal{B}(\pi^0\to e^+e^-\gamma(\gamma))=1.1836(6)\,\%$.
\end{abstract}

%\keywords{Decays of $\pi$ mesons; Electromagnetic form factors; Electromagnetic decays}

\maketitle

%=====================================================================================================
%*Introduction****************************************************************************************
%=====================================================================================================

\section{Introduction}
\label{sec:intro}

The present 3\% experimental precision on $\mathcal{B}(\pi^0_\text{D})$\footnote{%
We use the shorthand notation $\pi_f^0\coloneqq\pi^0\to f$, with $\text{D}\coloneqq e^+e^-\gamma$ and $\text{DD}\coloneqq e^+e^-e^+e^-$.
}~\cite{Tanabashi:2018oca} represents a limitation for rare-$\pi^0$-decay measurements, which commonly use the Dalitz decay $\pi^0_\text{D}$ for normalization, and is also becoming a limiting factor for rare-kaon-decay measurements.
An example is the $K^+\to\pi^+e^+e^-$ decay~\cite{Batley:2009aa}: accurate knowledge of $\mathcal{B}(\pi^0_\text{D})$ would improve the precision on the rate measurement by 30\,\%, and the precision on the low-energy parameter $a_+$~\cite{DAmbrosio:1998gur} by 10\,\%.
The uncertainty on $\mathcal{B}(\pi^0_\text{D})$ also dominates the precision on the $K^\pm\to\pi^\pm\pi^0e^+e^-$ rate measurement~\cite{Batley:2018hxd}, and is among the principal contributions to the uncertainties on the measured $K_\text{L,S}\to\pi^+\pi^-e^+e^-$ rates~\cite{Lai:2003ad}.
In these circumstances, considering the improving precision on rare-decay measurements, and the recent progress on the $\pi^0$-form-factor measurement~\cite{TheNA62:2016fhr} and radiative corrections for the $\pi^0_\text{D}$ decay~\cite{Husek:2015sma}, both a precision measurement of $\mathcal{B}(\pi^0_\text{D})$ and an updated theoretical evaluation of this quantity are becoming more important.

Branching ratios can serve to translate lifetimes into decay widths and vice versa.
There are several methods to determine the $\pi^0$ lifetime: a direct average-distance measurement of the highly-relativistic pion, the conserved-vector-current hypothesis connecting the vector form factor (i.e.\ charged pions) to the $\pi^0$ lifetime~\cite{Bychkov:2008ws}, and the Primakoff effect~\cite{Pirmakoff:1951pj}.
Since 2012 its Particle Data Group (PDG) value settled to $\tau_{\pi^0}^\text{PDG}=8.52(18)\times10^{-17}$\,s~\cite{Beringer:1900zz}.
Presently, the most precise $\pi^0$-lifetime measurements are given by two different methods: $\tau_{\pi^0}^\text{PrimEx}=8.32(23)\times10^{-17}$\,s~\cite{Larin:2010kq} (Primakoff effect, PrimEx experiment at JLab) and $\tau_{\pi^0}^\text{CERN}=8.97(28)\times10^{-17}$\,s~\cite{Atherton:1985av} (direct measurement, CERN).
It is clear that the situation is unsatisfactory and a new independent measurement is desirable.
For the Primakoff-effect-type $\pi^0$-lifetime measurements, $\mathcal{B}(\pi^0_{2\gamma})$ constitutes an essential input.

In this work we discuss the theoretical determination of the following ratio of decay widths:
\begin{equation}
R
\equiv\frac{\Gamma(\pi^0\to e^+e^-\gamma)}{\Gamma(\pi^0\to\gamma\gamma)}
=\frac{\mathcal{B}(\pi^0_\text{D})}{\mathcal{B}(\pi^0_{2\gamma})}\,.
\label{eq:R}
\end{equation}
The current PDG value $R|_\text{PDG}=11.88(35)\times10^{-3}$ is an average of experimental results, the most recent of which comes from 2008 and is based on archived ALEPH data~\cite{Beddall:2008zza}.
Other measurements with competitive uncertainties date back to 1981~\cite{Schardt:1980qd} and 1961~\cite{Samios:1961zz}.
The branching ratios $\mathcal{B}(\pi^0_{2\gamma})|_\text{PDG}=98.823(34)\,\%$ and $\mathcal{B}(\pi^0_\text{D})|_\text{PDG}=1.174(35)\,\%$~\cite{Tanabashi:2018oca} are subsequently obtained from a constrained fit.

Besides the direct extraction of $R$ from experiment, the shape of the singly-virtual $\pi^0$ transition form factor (TFF) can be measured.
This can be expanded in the transferred momentum squared, with the linear coefficient called the (TFF) slope $a_\pi$.
Since the slope embodies the most relevant input to the ratio $R$ regarding the (non-perturbative) low-energy QCD sector (the peculiar~\cite{Kampf:2009tk} TFF normalization $\mathcal{F}(0)$ conveniently drops out), its knowledge from experiment is crucial to obtain a model-independent prediction of $R$.
Recently, it was measured in the NA62 experiment, which analyzed 1.1$\times10^6$ fully reconstructed $\pi_\text{D}^0$ decays with the result $a_\pi^\text{NA62}=3.68(57)\,\%$~\cite{TheNA62:2016fhr}, taking into account the complete set of next-to-leading-order (NLO) radiative corrections in the QED sector~\cite{Husek:2015sma}.
The current PDG value is dominated by two inputs: the above NA62 result and the value provided by the CELLO Collaboration ($a_\pi^\text{CELLO}=3.26(37)\,\%$)~\cite{Behrend:1990sr} by (model-dependent) extrapolation from the space-like region.

Our calculation combines a wide range of theoretical models and available experimental results on the TFF shape and the well-established QED calculation including the complete set of NLO corrections, taking into account higher orders in the QED expansion in a conservative uncertainty estimate.
As such, it represents a precise and reliable improvement (by two orders of magnitude) to the current PDG-based value of $R$, which might be further used in various theoretical predictions and experimental analyses.
Moreover, for the first time, the slope corrections were not neglected in the bremsstrahlung contribution.
Finally, we present $R$ for the full as well as partial kinematic regions.

Measurements of the TFF shape or the ratio $R$ require significant theoretical input and depend crucially on the proper incorporation of radiative corrections.
Consequently, a statement that experiment itself provides a more relevant value of $R$ than our theoretical prediction is by its nature imprecise.
However, the computation would not be possible without the experimental evidence that the TFF slope lies within a certain range of values.
An example of how theoretical inputs influence the experimental values in this sector is the well-known discrepancy in the rare decay $\pi^0_{e^+e^-}$ driven most probably by the approximate radiative corrections~\cite{Bergstrom:1982wk} which do not agree with the exact calculation~\cite{Vasko:2011pi}; for details and discussion see Refs.~\cite{Husek:2014tna,Husek:2015wta}.

%=====================================================================================================
%*Theoretical framework****************************************************************************************
%=====================================================================================================

\section{Theoretical framework}
\label{sec:framework}

Considering the QED expansion, the leading-order (LO) $\pi^0$-Dalitz-decay differential width reads~\cite{Mikaelian:1972yg,Kampf:2005tz,Husek:2015sma}
\begin{equation}
\begin{split}
&\frac{\diff^2\Gamma^\text{LO}(x,y)}{\diff x\diff y}\\
&=\frac{\alpha}{\pi}\,\Gamma(\pi^0\to\gamma\gamma)\bigg|\frac{\mathcal{F}(M_\pi^2x)}{\mathcal{F}(0)}\bigg|^2\frac{(1-x)^3}{4x}\left[1+y^2+\frac{4m_e^2}{M_\pi^2x}\right],
\end{split}
\label{eq:dLOxy}
\end{equation}
where the two-photon decay width is parametrized as
\begin{equation}
\Gamma(\pi^0\to\gamma\gamma)
\equiv\frac{e^4M_\pi^3}{64\pi}|\mathcal{F}(0)|^2\,.
\label{eq:G2g}
\end{equation}
Above, $M_\pi$ and $m_e$ are the neutral-pion and electron masses, respectively.
The definition (\ref{eq:G2g}) holds to all orders in the QED and Chiral Perturbation Theory expansions~\cite{Kampf:2009tk} and covers also possible physics from beyond the Standard Model, simply putting these nontrivial dynamical effects into the TFF normalization $\mathcal{F}(0)$.
As usual, kinematical variables $x$ and $y$ are defined as
\begin{equation}
x=\frac{(p_{e^-}+p_{e^+})^2}{M_\pi^2}\,,\quad y=-\frac{2}{M_\pi^2}\frac{p_{\pi^0}\cdot(p_{e^-}-p_{e^+})}{(1-x)}\,,
\label{eq:defxy}
\end{equation}
with $p$ denoting four-momenta of respective particles.

\begin{table}[t]
\begin{ruledtabular}
{\scriptsize
\begin{tabular}{c | c c c c c c c}
source & VMD & LMD & THS & dispers. & Pad\'e aps.\ & NA62 & PDG\\
\hline
$a_\pi\,[\%]$ & 3.00 & 2.45 & 2.92(4) & 3.15(9) & 3.21(19) & 3.68(57) & 3.35(31)\\
$b_\pi\,[10^{-3}]$ & 0.90 & 0.74 & 0.87(2) & 1.14(4) & 1.04(22) & $\times$ & $\times$\\
\end{tabular}}
\end{ruledtabular}
\caption{
The slope and curvature of the singly-virtual pion TFF in various approaches.
The values given by the VMD, LMD and THS models are compared with the results of the recent dispersive calculation~\cite{Hoferichter:2018dmo,Hoferichter:2018kwz} incorporating inputs from both the space- and time-like regions (and updating Ref.~\cite{Hoferichter:2014vra}), the method of Pad\'e approximants~\cite{Masjuan:2017tvw} mainly based on the extrapolation of the space-like data (as was the previous work~\cite{Masjuan:2012wy}) and supported by the low-energy time-like data, the recent measurement performed by the NA62 experiment~\cite{TheNA62:2016fhr} or the PDG average~\cite{Tanabashi:2018oca}.
Inherent model uncertainties (due to large-$N_\text{c}$ and chiral limits) are not fully included in the THS value~\cite{Husek:2015wta}.
Additionally, a recent time-like-region measurement by the A2 Collaboration reads $a_\pi^\text{A2}=3.0(1.0)\,\%$~\cite{Adlarson:2016ykr}.
}
\label{tab:slope}
\end{table}

The slope $a_\pi$ and curvature $b_\pi$ of the singly-virtual pion TFF are defined in terms of the Taylor expansion in the invariant mass of the vector current~\cite{Berman:1960zz,Landsberg:1986fd}:
\begin{equation}
\bigg|\frac{\mathcal{F}(M_\pi^2x)}{\mathcal{F}(0)}\bigg|
\equiv f(x)
=1+a_\pi x+b_\pi x^2+\mathcal{O}(x^3)\,.
\label{eq:fx}
\end{equation}
This parametrization is sufficient in the whole (small) region relevant to the $\pi^0_\text{D}$ decay.
Having particular theoretical models at hand one can immediately explore the properties of $\mathcal{F}(q^2)$ and calculate $a_\pi$ and $b_\pi$.
As examples we briefly mention the vector-meson-dominance (VMD) ansatz~\cite{sakuraiVMD,Landsberg:1986fd} together with the lowest-meson-dominance (LMD)~\cite{Peris:1998nj,Knecht:1999gb} and two-hadron-saturation (THS)~\cite{Husek:2015wta} models; see Ref.~\cite{FFmodels} for more details.
These belong to a family of large-$N_\text{c}$ motivated analytic resonance-saturation models and as such they can be straightforwardly used in the calculation of radiative corrections.
By means of the first and second derivatives it is easy to find the analytic expressions for $a_\pi$ and $b_\pi$ within these models; for details see Section 5 of Ref.~\cite{Husek:2015wta}.
Numerical results are shown in Table~\ref{tab:slope}, together with other theoretical approaches and experimental results.

From Eq.~(\ref{eq:fx}) it follows that
\begin{equation}
f^2(x)
=1+2a_\pi x+(a_\pi^2+2b_\pi)x^2+\mathcal{O}(x^3)\,.
\label{eq:f2x}
\end{equation}
We can use the expansion (\ref{eq:f2x}) to obtain a simple formula for the LO width.
Inserting Eq.~(\ref{eq:f2x}) into Eq.~(\ref{eq:dLOxy}) and taking into account that $x\in(4m_e^2/M_\pi^2,1)$ and $y\in\left(-\beta(M_\pi^2x),\beta(M_\pi^2x)\right)$ with $\beta(s)\equiv\sqrt{1-{4m_e^2}/s}$, we get
\begin{equation}
\begin{split}
R^\text{LO}
=\frac{\alpha}{\pi}
\bigg[
&\frac43\ln\frac{M_\pi}{m_e}
-\frac13(7-a_\pi)
+\frac1{30}(a_\pi^2+2b_\pi)\\
&+(12-8a_\pi)\frac{m_e^2}{M_\pi^2}
+\mathcal{O}\bigg(\frac{m_e^4}{M_\pi^4}\bigg)
\bigg]\,.
\end{split}
\label{eq:RLO}
\end{equation}
This expression is a very good approximation with the precision $\eta(R^\text{LO})\simeq10^{-9}$ (evaluated for parameters close to the physical ones).
As is a common practice, e.g.\ in Ref.~\cite{TheNA62:2016fhr}, further on the prescription $f^2(x)\big|_\text{lin.}\equiv(1+a_\pi x)^2$ is used, which is accurate to first order in $x$.
The effect of the missing $2b_\pi x^2$ term is negligible, since the region where such a difference arises (i.e.\ when $x\simeq1$) is suppressed; cf.\ Eq.~(\ref{eq:dLOxy}) and also Table~\ref{tab:xcut} later on.
If we consider $b_\pi\simeq a_\pi^2$, as suggested by the models (cf.\ Table~\ref{tab:slope}), we introduce an error of $\sigma(a_\pi)\simeq a_\pi^2/5$, i.e.\ a relative error $\eta(a_\pi)\lesssim1$\,\%, on the estimate of $a_\pi$ being well under the current experimental precision.
The previous discussion also implies that the effect of the $a_\pi x$ term on the Dalitz-decay rate is limited, letting us provide a very precise determination of $R$ while allowing for 20\,\% uncertainty on $a_\pi$.
Finally, let us note that dropping $b_\pi$ out of Eq.~(\ref{eq:RLO}) decreases its precision down to $\eta(R^\text{LO}|_{b_\pi=0})\simeq10^{-5}$, being still a good approximation for our purpose in view of the above discussion.

\begin{figure}[t]
\includegraphics[width=\columnwidth]{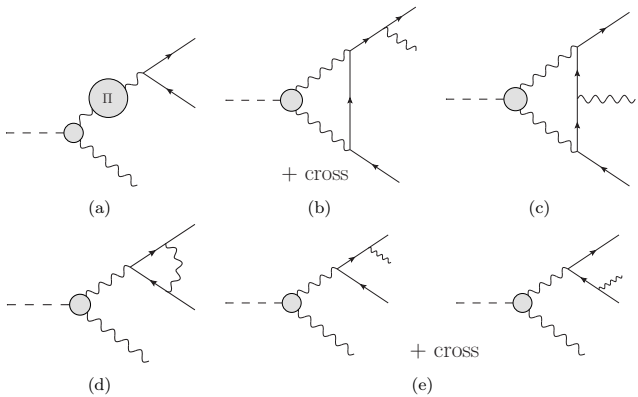}
\caption{
\label{fig:diagrams}
NLO QED radiative corrections for $\pi^0_\text{D}$: (a)~vacuum-polarization insertion; (b),(c)~one-loop 1$\gamma$IR contribution; (d)~vertex correction; (e)~bremsstrahlung.
}
\end{figure}

In the rest of the section we address the NLO QED-sector radiative corrections; see Fig.~\ref{fig:diagrams} for Feynman diagrams.
It is convenient to introduce the NLO correction $\delta$ to the LO differential width (and thus to write schematically $\text{d}\Gamma=(1+\delta+\ldots)\,\text{d}\Gamma^\text{LO}$), which can be in general (for the two- and one-fold differential case, respectively) defined as
\begin{equation}
\delta(x,y)
=\frac{\diff^2\Gamma^\text{NLO}}{\diff x\diff y}\bigg/\frac{\diff^2\Gamma^\text{LO}}{\diff x\diff y}\,,\quad
\delta(x)
=\frac{\diff\Gamma^\text{NLO}}{\diff x}\bigg/\frac{\diff\Gamma^\text{LO}}{\diff x}\,.
\label{eq:dxy}
\end{equation}
One can obtain $\delta(x)$ from $\delta(x,y)$ using the following prescription:
\begin{equation}
\begin{split}
\delta(x)
&=\frac38\frac{1}{\beta(M_\pi^2x)}\left(1+\frac{2m_e^2}{M_\pi^2x}\right)^{-1}\\
&\times\int_{-\beta(M_\pi^2x)}^{\beta(M_\pi^2x)}\delta(x,y)\left[1+y^2+\frac{4m_e^2}{M_\pi^2x}\right]\diff y\,.
\end{split}
\label{eq:dx}
\end{equation}
To calculate the NLO radiative corrections we use the approach documented in Refs.~\cite{Husek:2015sma,Husek:2017vmo}, which reviewed and extended the classical work of Mikaelian and Smith~\cite{Mikaelian:1972yg}.
Hence, together with the bremsstrahlung (BS) beyond the soft-photon approximation, we take into account in the following calculations the one-photon-irreducible (1$\gamma$IR) contribution; see Figs.~\ref{fig:diagrams}(b) and \ref{fig:diagrams}(c).

For historical reasons~\cite{Mikaelian:1972yg,Tupper:1983uw,Lambin:1985sb,Tupper:1986yk,Kampf:2005tz}, let us discuss the case when the 1$\gamma$IR contribution to the NLO radiative corrections is not considered in the analysis to extract the TFF slope from the data.
If we start with the equation among the measured spectral shapes (one-fold differential widths) and eliminate ${\diff\Gamma^\text{LO}}/{\diff x}|_{f(x)=1}$ from both sides, take the expansion (\ref{eq:f2x}) to the linear order, and neglect the corrections of order $\alpha a_\pi$, we find
\begin{equation}
\Delta a_\pi x
\equiv\left(a_\pi-a_\pi^{1\gamma\text{IR}}\right)x
\simeq-\frac12\delta_{1\gamma\text{IR}}^\text{NLO}(x)\,,\quad x\ll1\,.
\end{equation}
Numerically, $\Delta a_\pi\simeq0.5\,\%$~\cite{Kampf:2005tz}.
This is the value to be {\em added} to the experimental value $a_\pi^{1\gamma\text{IR}}$ (extracted neglecting the 1$\gamma$IR contribution) in order to find an estimate of the pure-low-energy-QCD-sector parameter $a_\pi$ with all the QED radiative corrections subtracted.
Above, $\delta_{1\gamma\text{IR}}^\text{NLO}(x)$ is calculated from $\delta^{1\gamma\text{IR}}(x,y)$ stated in Section IV of Ref.~\cite{Husek:2015sma} using the prescription (\ref{eq:dx}).
Note that the 1$\gamma$IR contribution was already taken into account in the NA62 analysis~\cite{TheNA62:2016fhr} and $a_\pi^\text{NA62}$ does not need to be corrected by $\Delta a_\pi$.

Finally, taking the prescription $f^2(x)\big|_\text{lin.}$ and the NLO QED radiative corrections to approximate the exact Dalitz-decay differential width beyond LO (and consequently $\Delta R\equiv R-R^\text{LO}$) we arrive at
\begin{equation}
\begin{split}
\Delta R
\simeq
R^\text{NLO}
&\equiv\frac{\alpha}{\pi}
\iint
\,(1+a_\pi x)^2
\delta(x,y)\\
&\times\frac{(1-x)^3}{4x}\left[1+y^2+\frac{4m_e^2}{M_\pi^2x}\right]\diff x\diff y\,.
\label{eq:RatNLO}
\end{split}
\end{equation}

%=====================================================================================================
%*Calculation****************************************************************************************
%=====================================================================================================

\section{Calculation and uncertainty estimation}
\label{sec:calculation}

Our aim now is to precisely and reliably (using conservative error estimates) determine $R$.
In the following we choose $a_\pi^\text{univ}\equiv3.55(70)\,\%$ by stretching the uncertainty band over the whole interval of values suggested by different approaches; cf.\ Table~\ref{tab:slope}.
From Eq.~(\ref{eq:RLO}) we get
\begin{equation}
R^\text{LO}
=11.879(5)\times10^{-3}
\label{eq:RLOnum}
\end{equation}
and based on (\ref{eq:RatNLO}) we arrive at
\begin{equation}
\Delta R
=0.099(3)\times10^{-3}\,.
\label{eq:dR}
\end{equation}
During the estimation of the above uncertainty, higher orders in the QED expansion were considered, surpassing in size the uncertainty stemming from the TFF dependence.
In this regard, we can take the absolute value of the {\em dominant} correction ($R_\text{BS,div}^\text{NLO}$; see Table~\ref{tab:corr}) as the typical expected maximum value appearing in NLO and anticipate that the NNLO correction is suppressed compared to the NLO one in the similar manner as NLO is suppressed with respect to LO: circa on the level of 3\,\%.
\begin{table}[t]
\begin{ruledtabular}
\begin{tabular}{c c c c | c}
$R_\text{virt}^\text{NLO}$ & $R_\text{BS,conv}^\text{NLO}$ & $R_\text{BS,div}^\text{NLO}$ & $R_{1\gamma\text{IR}}^\text{NLO}$ & $R^\text{NLO}$ \\
\hline
$-$0.0750(2) & $-$0.15759(2) & 0.3363(3) & $-$0.00466(2) & 0.09911(7)\\
\end{tabular}
\end{ruledtabular}
\caption{
Individual contributions of the NLO radiative corrections for $R$ in [$10^{-3}$].
The `virt' label stands for virtual corrections (Figs.~\ref{fig:diagrams}(a) and \ref{fig:diagrams}(d)) and `div' (`conv') label the divergent (convergent) parts of the bremsstrahlung contribution (Fig.~\ref{fig:diagrams}(e))~\cite{details}.
The listed uncertainties stem from the uncertainty of $a_\pi^\text{univ}$.
In the case of the 1$\gamma$IR correction (Figs.~\ref{fig:diagrams}(b) and \ref{fig:diagrams}(c)), a particular model for the doubly-virtual TFF (LMD, etc.) is necessary to introduce.
The resulting model dependence is suppressed~\cite{Husek:2015sma} and related uncertainty included.
}
\label{tab:corr}
\end{table}
This uncertainty is already conservative: the total NLO correction is on the level of 1\,\%.
Summing Eqs.~(\ref{eq:RLOnum}) and (\ref{eq:dR}) we finally obtain
\begin{equation}
R
=11.978(5)(3)\times10^{-3}\,.
\label{eq:Rall}
\end{equation}
This is one of the main results of the presented work.
The former uncertainty stands for the TFF effects and the latter for neglecting the higher-order corrections;%
\footnote{
Relaxing the requirement of providing a conservative value, one can significantly reduce the former uncertainty (stemming from the TFF effects) taking into account a particular result from Table~\ref{tab:slope}: e.g.\ with the most precise entry --- the dispersion-theoretical result~\cite{Hoferichter:2018dmo,Hoferichter:2018kwz} --- by factor of 8.
Higher-order QED corrections would need to be computed to achieve an additional gain of precision.
}
$m_e$, $M_\pi$ and $\alpha$ are known very precisely.

This calculation also includes all contributions from the decays where additional photon(s) with arbitrarily high (kinematically allowed) energies are radiated.
Indeed, the bremsstrahlung correction at NLO (calculated {\em \`a la} Refs.~\cite{Mikaelian:1972yg,Husek:2015sma}) takes into account an additional final-state photon and integrates over its energy and emission angle without any additional cuts.
The results are thus meant to be used for the {\em inclusive} process.
However, quantities for {\em exclusive} processes can be obtained in a similar way while introducing some specific cut-off in the bremsstrahlung correction $\delta^\text{BS}(x,y)$.
A combined approach was used in the analysis of the recent NA62 measurement~\cite{TheNA62:2016fhr}, when an additional photon was simulated above the cut-off given by the detector sensitivity.
To conclude, for each experimental setup the specific approach for including radiative corrections must be used.
When it applies, we explicitly state (as in the abstract) that the results include an additional final-state photon, denoting it as ($\gamma$).
We also take this tacitly into account in the results for $R$, e.g.\ in Eq.~(\ref{eq:Rall}).

In experiments, specific kinematic regions might be considered.
The sample values for
\begin{equation}
R(x_\text{cut})
\equiv\frac{\mathcal{B}(\pi^0\to e^+e^-\gamma(\gamma),x>x_\text{cut})}{\mathcal{B}(\pi^0_{2\gamma})}
\end{equation}
are listed in Table~\ref{tab:xcut}, using which one can also obtain values for any intermediate region.
%%
%\begin{table*}[t]
%\begin{ruledtabular}
%{\scriptsize
%\begin{tabular}{c | c c c c c c}
%$x_\text{cut}$ & 0.05 & 0.10 & 0.15 & 0.20 & 0.25 & \\
%$R(x_\text{cut})\,[10^{-5}]$ & 203.72(45) & 117.03(36) & 74.38(29) & 49.11(23) & 32.92(17) \\
%\hline
%$x_\text{cut}$ & 0.30 & 0.35 & 0.40 & 0.45 & 0.50 \\
%$R(x_\text{cut})\,[10^{-5}]$ & 22.13(13) & 14.787(97) & 9.756(71) & 6.313(50) & 3.978(34) \\
%\hline
%$x_\text{cut}$ & 0.55 & 0.60 & 0.65 & 0.70 & 0.75 \\
%$R(x_\text{cut})\,[10^{-6}]$ & 24.20(22) & 14.07(14) & 7.703(80) & 3.890(43) & 1.757(21) \\
%\hline
%$x_\text{cut}$ & 0.80 & 0.85 & 0.90 & 0.95 & 1.00 \\
%$R(x_\text{cut})\,[10^{-8}]$ & 67.28(85) & 19.81(27) & 3.594(54) & 0.1967(35) & 0 \\
%%
%\end{tabular}}
%\end{ruledtabular}
%\caption{
%The values of $R(x_\text{cut})$ for chosen sample values of $x_\text{cut}$.
%To be suitable for interpolation, higher precision is used.
%The quoted uncertainties are dominated by the TFF-slope knowledge (for its value we assume $a_\pi^\text{univ}$); the additional 3\% uncertainty covering the higher-order corrections is also included.
%Note different additional multiplicative factors depending on $x_\text{cut}$.
%}
%\label{tab:xcut}
%\end{table*}
%%
%
\begin{table*}[bt]
\begin{ruledtabular}
\begin{tabular}{c r c | c r c | c r c | c r}
$x_\text{cut}$ & $R(x_\text{cut})\,[10^{-5}]$ && $x_\text{cut}$ & $R(x_\text{cut})\,[10^{-5}]$ && $x_\text{cut}$ & $R(x_\text{cut})\,[10^{-6}]$ && $x_\text{cut}$ & $R(x_\text{cut})\,[10^{-8}]$ \\
\hline
0.05 & 203.72(45) && 0.30 & 22.13(13)~\, && 0.55 & 24.20(22)~\,& & 0.80 & 67.28(85)~~\; \\
0.10 & 117.03(36) && 0.35 & 14.787(97) && 0.60 & 14.07(14)~\, && 0.85 & 19.81(27)~~\; \\
0.15 & 74.38(29) && 0.40 & 9.756(71) && 0.65 & 7.703(80) && 0.90 & 3.594(54)~\, \\
0.20 & 49.11(23) && 0.45 & 6.313(50) && 0.70 & 3.890(43) && 0.95 & 0.1967(35) \\
0.25 & 32.92(17) && 0.50 & 3.978(34) && 0.75 & 1.757(21) && 1.00 & 0\hspace{1.32cm} \\
\end{tabular}
\end{ruledtabular}
\caption{
The values of $R(x_\text{cut})$ for chosen sample values of $x_\text{cut}$.
To be suitable for interpolation, higher precision is used.
The quoted uncertainties are dominated by the TFF-slope knowledge (for its value we assume $a_\pi^\text{univ}$); the additional 3\% uncertainty covering the higher-order corrections is also included.
Note different additional multiplicative factors depending on $x_\text{cut}$.
}
\label{tab:xcut}
\end{table*}
As an example, in the $\pi^0$-rare-decay measurement performed by KTeV~\cite{Abouzaid:2006kk} the region $x>x_\text{cut}=0.232$ was used for the Dalitz decay, which served as the normalization channel in this search.
The direct calculation based on this work leads to $R(0.232)=0.380(2)\times10^{-3}$ and the interpolation based on Table~\ref{tab:xcut} gives $R(0.232)|_\text{intpol.}=0.379(2)\times10^{-3}$, which is compatible within uncertainties.
In Ref.~\cite{Niclasen:2006tz} the value $[R(0.2319)/R]|_\text{KTeV}=0.0319$ was used, which is compatible with our calculation: $R(0.232)/R=0.0317(2)$.

Based on $R$, we can predict $\mathcal{B}(\pi^0_{2\gamma})$ and $\mathcal{B}(\pi^0_\text{D})$.
Considering the uncertainty of $R$, we can write
\begin{equation}
\begin{split}
&1-\mathcal{B}(\pi^0_\text{DD})
\simeq
\mathcal{B}(\pi^0_{2\gamma})+\mathcal{B}(\pi^0_{\text{D}(\gamma)})\,,
\end{split}
\label{eq:1eqB}
\end{equation}
since the branching ratios of other decay modes are smaller than 10$^{-6}$.
Using $\mathcal{B}(\pi^0_\text{DD})=3.3(2)\times10^{-5}$~\cite{Tanabashi:2018oca,Kampf:2018wau} (double-Dalitz decay), we find
\begin{equation}
\mathcal{B}(\pi^0_{2\gamma})
\simeq
\frac{1-\mathcal{B}(\pi^0_\text{DD})}{1+R}
=98.8131(6)\,\%\,.
\label{eq:Bgg}
\end{equation}
Note that taking tacitly into account inclusive Dalitz decays in Eq.~(\ref{eq:1eqB}) is justified and contributes to the relevant decay modes.
Finally, the Dalitz-decay branching ratio reads
\begin{equation}
\begin{split}
&\mathcal{B}(\pi^0_{\text{D}(\gamma)})
\simeq
\frac{R}{1+R}\,[1-\mathcal{B}(\pi^0_\text{DD})]
=1.1836(6)\,\%\,.
\end{split}
\label{eq:BDalitz}
\end{equation}
The above results are compatible with the PDG averages, exhibiting much higher precision.

Let us see how the new result on the Dalitz-decay branching ratio (\ref{eq:BDalitz}) influences a completely different family of processes on a simple example of the $K^+\to\pi^+ e^+e^-$ decay measurements.
The low-energy parameters $a_+$ and $b_+$ were measured by the NA48/2 Collaboration to be $a_+=-0.578(16)$ and $b_+=-0.779(66)$, leading to the model-dependent branching ratio $\mathcal{B}(K^+\to\pi^+ e^+e^-)=3.11(12)\times10^{-7}$, using the 2008 PDG average $\mathcal{B}(\pi^0_\text{D})=1.198(32)\,\%$~\cite{Amsler:2008zzb} for normalization~\cite{Batley:2009aa}.
The central value of our result (\ref{eq:BDalitz}) is 1.2\,\% lower than the quoted PDG average and has a negligible error.
The remaining external uncertainty on the measurement~\cite{Batley:2009aa} related to the normalization comes from $\mathcal{B}(K^+\to\pi^+\pi^0)$ known to 0.4\% precision.
The corrected values are $a_+=-0.575(14)$, $b_+=-0.771(64)$ and $\mathcal{B}(K^+\to\pi^+ e^+e^-)=3.07(10)\times10^{-7}$.
Note that considering the external errors on $a_+$ and $b_+$ quoted in Ref.~\cite{Batley:2009aa}, further experimental progress on $K^+\to\pi^+ e^+e^-$ measurement would be impossible without improvement on $\mathcal{B}(\pi^0_\text{D})$.

%=====================================================================================================
%*Comparison and conclusion****************************************************************************************
%=====================================================================================================

\section{Comparison and conclusion}
\label{sec:comparison}

Radiative corrections for the integral decay width were first addressed by Joseph~\cite{Joseph:1960zz}, who numerically arrived to $\Delta R|_\text{Jph.}=0.105\times10^{-3}$ neglecting, among others, the pion TFF slope.
A simple analytical prescription in the limit of the vanishing electron mass was later found by Lautrup and Smith~\cite{Lautrup:1971ew}:
\begin{equation}
\begin{split}
\Delta R\big|_\text{L\&S}
&=\left(\frac{\alpha}{\pi}\right)^2
\left[
\frac89\ln^2
%\left(
\frac{M_\pi}{m_e}
%\right)
-\frac19\left(19-4a_\pi\right)\ln
%\left(
\frac{M_\pi}{m_e}
%\right)
\right.\\
&+\left.2\zeta(3)
-\frac2{27}\pi^2
+\frac{137}{81}
-\frac{63}{108}a_\pi
+\mathcal{O}\bigg(\frac{m_e}{M_\pi}\bigg)
\right].
\end{split}
\end{equation}
Numerically, $\Delta R\big|_\text{L\&S}^{a_\pi=0}=0.10378\times10^{-3}$ and $\Delta R\big|_\text{L\&S}^{a_\pi^\text{univ}}=0.10414(7)\times10^{-3}$.
The two approaches are compatible and should be compared with our result (\ref{eq:dR}).
However, the 1$\gamma$IR contribution was, due to inappropriate assumptions and arguments based on Low's theorem~\cite{Low:1958sn,Adler:1966gc,Pestleau:1967snm}, considered negligible
and left out; see also Refs.~\cite{Mikaelian:1972yg,Lambin:1985sb}.
The exact calculation shows its significance~\cite{Tupper:1983uw,Tupper:1986yk,Kampf:2005tz,Husek:2015sma} and it thus embodies the main source of the difference between our result and the previous works.
Moreover, the symmetrization with respect to the two photons in the bremsstrahlung contribution was neglected in Refs.~\cite{Joseph:1960zz,Lautrup:1971ew}.
This interference of the diagrams from Fig.~\ref{fig:diagrams}(e) is indeed negligible and corresponds (for $a_\pi=0$) to $\Delta R_\text{interf}^\text{BS}=0.000360\times10^{-3}$.
Let us stress again that the prediction (\ref{eq:Rall}) is based on the complete calculation which includes the entire bremsstrahlung and 1$\gamma$IR contributions.
Here, TFF effects were taken into account also in the bremsstrahlung correction~\cite{details} and the mass of the final-state leptons was {\em not} neglected.

Our main result (\ref{eq:Rall}) together with the value (\ref{eq:Bgg}) should be considered as an alternative to the current PDG averages which opens the way to a new level of precision for a whole class of other processes, for instance for the already mentioned kaon decays.
Similarly, the current situation, when the precision on $\mathcal{B}(\pi^0_\text{D})|_\text{PDG}$ dominates the uncertainty on $\mathcal{B}(\pi^0_\text{DD})$~\cite{Abouzaid:2008cd} and is the largest source of uncertainty on $\mathcal{B}(\pi^0_{e^+e^-})$~\cite{Abouzaid:2006kk}, is improved.

%=====================================================================================================
%*Acknowledgment****************************************************************************************
%=====================================================================================================

\begin{acknowledgments}
We thank G.\ D'Ambrosio, M.\ Hoferichter and A.\ Portelli for initial suggestions, P. Sanchez-Puertas for helpful discussions and J. Portol\'es for comments on the manuscript.

This work has been supported in part by
the Agencia Estatal de Investigaci\'on (AEI, ES) and the European Regional Development Fund (ERDF, EU) [Grants No.\ FPA2014-53631-C2-1-P, FPA2017-84445-P and SEV-2014-0398],
by Generalitat Valenciana [Grant No.\ PROMETEO/2017/053],
by the Czech Science Foundation grant GA\v{C}R 18-17224S and by the ERC starting grant 336581 ``KaonLepton''.
\end{acknowledgments}

%\bibliography{refs}

\begin{thebibliography}{56}%
\makeatletter
\providecommand \@ifxundefined [1]{%
 \@ifx{#1\undefined}
}%
\providecommand \@ifnum [1]{%
 \ifnum #1\expandafter \@firstoftwo
 \else \expandafter \@secondoftwo
 \fi
}%
\providecommand \@ifx [1]{%
 \ifx #1\expandafter \@firstoftwo
 \else \expandafter \@secondoftwo
 \fi
}%
\providecommand \natexlab [1]{#1}%
\providecommand \enquote  [1]{``#1''}%
\providecommand \bibnamefont  [1]{#1}%
\providecommand \bibfnamefont [1]{#1}%
\providecommand \citenamefont [1]{#1}%
\providecommand \href@noop [0]{\@secondoftwo}%
\providecommand \href [0]{\begingroup \@sanitize@url \@href}%
\providecommand \@href[1]{\@@startlink{#1}\@@href}%
\providecommand \@@href[1]{\endgroup#1\@@endlink}%
\providecommand \@sanitize@url [0]{\catcode `\\12\catcode `\$12\catcode
  `\&12\catcode `\#12\catcode `\^12\catcode `\_12\catcode `\%12\relax}%
\providecommand \@@startlink[1]{}%
\providecommand \@@endlink[0]{}%
\providecommand \url  [0]{\begingroup\@sanitize@url \@url }%
\providecommand \@url [1]{\endgroup\@href {#1}{\urlprefix }}%
\providecommand \urlprefix  [0]{URL }%
\providecommand \Eprint [0]{\href }%
\providecommand \doibase [0]{http://dx.doi.org/}%
\providecommand \selectlanguage [0]{\@gobble}%
\providecommand \bibinfo  [0]{\@secondoftwo}%
\providecommand \bibfield  [0]{\@secondoftwo}%
\providecommand \translation [1]{[#1]}%
\providecommand \BibitemOpen [0]{}%
\providecommand \bibitemStop [0]{}%
\providecommand \bibitemNoStop [0]{.\EOS\space}%
\providecommand \EOS [0]{\spacefactor3000\relax}%
\providecommand \BibitemShut  [1]{\csname bibitem#1\endcsname}%
\let\auto@bib@innerbib\@empty
%</preamble>
\bibitem [{\citenamefont {Tanabashi}\ \emph {et~al.}(2018)\citenamefont
  {Tanabashi} \emph {et~al.}}]{Tanabashi:2018oca}%
  \BibitemOpen
  \bibfield  {author} {\bibinfo {author} {\bibfnamefont {M.}~\bibnamefont
  {Tanabashi}} \emph {et~al.} (\bibinfo {collaboration} {Particle Data
  Group}),\ }\href {\doibase 10.1103/PhysRevD.98.030001} {\bibfield  {journal}
  {\bibinfo  {journal} {Phys. Rev.}\ }\textbf {\bibinfo {volume} {D98}},\
  \bibinfo {pages} {030001} (\bibinfo {year} {2018})}\BibitemShut {NoStop}%
%%CITATION = PHRVA,D98,030001;%%
\bibitem [{\citenamefont {Batley}\ \emph {et~al.}(2009)\citenamefont {Batley}
  \emph {et~al.}}]{Batley:2009aa}%
  \BibitemOpen
  \bibfield  {author} {\bibinfo {author} {\bibfnamefont {J.~R.}\ \bibnamefont
  {Batley}} \emph {et~al.} (\bibinfo {collaboration} {NA48/2}),\ }\href
  {\doibase 10.1016/j.physletb.2009.05.040} {\bibfield  {journal} {\bibinfo
  {journal} {Phys. Lett.}\ }\textbf {\bibinfo {volume} {B677}},\ \bibinfo
  {pages} {246} (\bibinfo {year} {2009})},\ \Eprint
  {http://arxiv.org/abs/0903.3130} {arXiv:0903.3130 [hep-ex]} \BibitemShut
  {NoStop}%
%%CITATION = ARXIV:0903.3130;%%
\bibitem [{\citenamefont {D'Ambrosio}\ \emph {et~al.}(1998)\citenamefont
  {D'Ambrosio}, \citenamefont {Ecker}, \citenamefont {Isidori},\ and\
  \citenamefont {Portoles}}]{DAmbrosio:1998gur}%
  \BibitemOpen
  \bibfield  {author} {\bibinfo {author} {\bibfnamefont {G.}~\bibnamefont
  {D'Ambrosio}}, \bibinfo {author} {\bibfnamefont {G.}~\bibnamefont {Ecker}},
  \bibinfo {author} {\bibfnamefont {G.}~\bibnamefont {Isidori}}, \ and\
  \bibinfo {author} {\bibfnamefont {J.}~\bibnamefont {Portoles}},\ }\href
  {\doibase 10.1088/1126-6708/1998/08/004} {\bibfield  {journal} {\bibinfo
  {journal} {JHEP}\ }\textbf {\bibinfo {volume} {08}},\ \bibinfo {pages} {004}
  (\bibinfo {year} {1998})},\ \Eprint {http://arxiv.org/abs/hep-ph/9808289}
  {arXiv:hep-ph/9808289 [hep-ph]} \BibitemShut {NoStop}%
%%CITATION = HEP-PH/9808289;%%
\bibitem [{\citenamefont {Batley}\ \emph {et~al.}(2019)\citenamefont {Batley}
  \emph {et~al.}}]{Batley:2018hxd}%
  \BibitemOpen
  \bibfield  {author} {\bibinfo {author} {\bibfnamefont {J.~R.}\ \bibnamefont
  {Batley}} \emph {et~al.} (\bibinfo {collaboration} {NA48/2}),\ }\href
  {\doibase 10.1016/j.physletb.2018.11.046} {\bibfield  {journal} {\bibinfo
  {journal} {Phys. Lett.}\ }\textbf {\bibinfo {volume} {B788}},\ \bibinfo
  {pages} {552} (\bibinfo {year} {2019})},\ \Eprint
  {http://arxiv.org/abs/1809.02873} {arXiv:1809.02873 [hep-ex]} \BibitemShut
  {NoStop}%
%%CITATION = ARXIV:1809.02873;%%
\bibitem [{\citenamefont {Lai}\ \emph {et~al.}(2003)\citenamefont {Lai} \emph
  {et~al.}}]{Lai:2003ad}%
  \BibitemOpen
  \bibfield  {author} {\bibinfo {author} {\bibfnamefont {A.}~\bibnamefont
  {Lai}} \emph {et~al.} (\bibinfo {collaboration} {NA48}),\ }\href {\doibase
  10.1140/epjc/s2003-01252-y} {\bibfield  {journal} {\bibinfo  {journal} {Eur.
  Phys. J.}\ }\textbf {\bibinfo {volume} {C30}},\ \bibinfo {pages} {33}
  (\bibinfo {year} {2003})}\BibitemShut {NoStop}%
%%CITATION = EPHJA,C30,33;%%
\bibitem [{\citenamefont {Lazzeroni}\ \emph {et~al.}(2017)\citenamefont
  {Lazzeroni} \emph {et~al.}}]{TheNA62:2016fhr}%
  \BibitemOpen
  \bibfield  {author} {\bibinfo {author} {\bibfnamefont {C.}~\bibnamefont
  {Lazzeroni}} \emph {et~al.} (\bibinfo {collaboration} {NA62}),\ }\href
  {\doibase 10.1016/j.physletb.2017.02.042} {\bibfield  {journal} {\bibinfo
  {journal} {Phys. Lett.}\ }\textbf {\bibinfo {volume} {B768}},\ \bibinfo
  {pages} {38} (\bibinfo {year} {2017})},\ \Eprint
  {http://arxiv.org/abs/1612.08162} {arXiv:1612.08162 [hep-ex]} \BibitemShut
  {NoStop}%
%%CITATION = ARXIV:1612.08162;%%
\bibitem [{\citenamefont {Husek}\ \emph {et~al.}(2015)\citenamefont {Husek},
  \citenamefont {Kampf},\ and\ \citenamefont {Novotn\'y}}]{Husek:2015sma}%
  \BibitemOpen
  \bibfield  {author} {\bibinfo {author} {\bibfnamefont {T.}~\bibnamefont
  {Husek}}, \bibinfo {author} {\bibfnamefont {K.}~\bibnamefont {Kampf}}, \ and\
  \bibinfo {author} {\bibfnamefont {J.}~\bibnamefont {Novotn\'y}},\ }\href
  {\doibase 10.1103/PhysRevD.92.054027} {\bibfield  {journal} {\bibinfo
  {journal} {Phys. Rev.}\ }\textbf {\bibinfo {volume} {D92}},\ \bibinfo {pages}
  {054027} (\bibinfo {year} {2015})},\ \Eprint
  {http://arxiv.org/abs/1504.06178} {arXiv:1504.06178 [hep-ph]} \BibitemShut
  {NoStop}%
%%CITATION = ARXIV:1504.06178;%%
\bibitem [{\citenamefont {Bychkov}\ \emph {et~al.}(2009)\citenamefont {Bychkov}
  \emph {et~al.}}]{Bychkov:2008ws}%
  \BibitemOpen
  \bibfield  {author} {\bibinfo {author} {\bibfnamefont {M.}~\bibnamefont
  {Bychkov}} \emph {et~al.},\ }\href {\doibase 10.1103/PhysRevLett.103.051802}
  {\bibfield  {journal} {\bibinfo  {journal} {Phys. Rev. Lett.}\ }\textbf
  {\bibinfo {volume} {103}},\ \bibinfo {pages} {051802} (\bibinfo {year}
  {2009})},\ \Eprint {http://arxiv.org/abs/0804.1815} {arXiv:0804.1815
  [hep-ex]} \BibitemShut {NoStop}%
%%CITATION = ARXIV:0804.1815;%%
\bibitem [{\citenamefont {Primakoff}(1951)}]{Pirmakoff:1951pj}%
  \BibitemOpen
  \bibfield  {author} {\bibinfo {author} {\bibfnamefont {H.}~\bibnamefont
  {Primakoff}},\ }\href {\doibase 10.1103/PhysRev.81.899} {\bibfield  {journal}
  {\bibinfo  {journal} {Phys. Rev.}\ }\textbf {\bibinfo {volume} {81}},\
  \bibinfo {pages} {899} (\bibinfo {year} {1951})}\BibitemShut {NoStop}%
%%CITATION = PHRVA,81,899;%%
\bibitem [{\citenamefont {Beringer}\ \emph {et~al.}(2012)\citenamefont
  {Beringer} \emph {et~al.}}]{Beringer:1900zz}%
  \BibitemOpen
  \bibfield  {author} {\bibinfo {author} {\bibfnamefont {J.}~\bibnamefont
  {Beringer}} \emph {et~al.} (\bibinfo {collaboration} {Particle Data Group}),\
  }\href {\doibase 10.1103/PhysRevD.86.010001} {\bibfield  {journal} {\bibinfo
  {journal} {Phys. Rev.}\ }\textbf {\bibinfo {volume} {D86}},\ \bibinfo {pages}
  {010001} (\bibinfo {year} {2012})}\BibitemShut {NoStop}%
%%CITATION = PHRVA,D86,010001;%%
\bibitem [{\citenamefont {Larin}\ \emph {et~al.}(2011)\citenamefont {Larin}
  \emph {et~al.}}]{Larin:2010kq}%
  \BibitemOpen
  \bibfield  {author} {\bibinfo {author} {\bibfnamefont {I.}~\bibnamefont
  {Larin}} \emph {et~al.} (\bibinfo {collaboration} {PrimEx}),\ }\href
  {\doibase 10.1103/PhysRevLett.106.162303} {\bibfield  {journal} {\bibinfo
  {journal} {Phys. Rev. Lett.}\ }\textbf {\bibinfo {volume} {106}},\ \bibinfo
  {pages} {162303} (\bibinfo {year} {2011})},\ \Eprint
  {http://arxiv.org/abs/1009.1681} {arXiv:1009.1681 [nucl-ex]} \BibitemShut
  {NoStop}%
%%CITATION = ARXIV:1009.1681;%%
\bibitem [{\citenamefont {Atherton}\ \emph {et~al.}(1985)\citenamefont
  {Atherton} \emph {et~al.}}]{Atherton:1985av}%
  \BibitemOpen
  \bibfield  {author} {\bibinfo {author} {\bibfnamefont {H.~W.}\ \bibnamefont
  {Atherton}} \emph {et~al.},\ }\href {\doibase 10.1016/0370-2693(85)90744-0}
  {\bibfield  {journal} {\bibinfo  {journal} {Phys. Lett.}\ }\textbf {\bibinfo
  {volume} {158B}},\ \bibinfo {pages} {81} (\bibinfo {year}
  {1985})}\BibitemShut {NoStop}%
%%CITATION = PHLTA,158B,81;%%
\bibitem [{\citenamefont {Beddall}\ and\ \citenamefont
  {Beddall}(2008)}]{Beddall:2008zza}%
  \BibitemOpen
  \bibfield  {author} {\bibinfo {author} {\bibfnamefont {A.}~\bibnamefont
  {Beddall}}\ and\ \bibinfo {author} {\bibfnamefont {A.}~\bibnamefont
  {Beddall}},\ }\href {\doibase 10.1140/epjc/s10052-008-0539-0} {\bibfield
  {journal} {\bibinfo  {journal} {Eur. Phys. J.}\ }\textbf {\bibinfo {volume}
  {C54}},\ \bibinfo {pages} {365} (\bibinfo {year} {2008})}\BibitemShut
  {NoStop}%
%%CITATION = EPHJA,C54,365;%%
\bibitem [{\citenamefont {Schardt}\ \emph {et~al.}(1981)\citenamefont
  {Schardt}, \citenamefont {Frank}, \citenamefont {Hoffman}, \citenamefont
  {Mischke}, \citenamefont {Moir},\ and\ \citenamefont
  {Thompson}}]{Schardt:1980qd}%
  \BibitemOpen
  \bibfield  {author} {\bibinfo {author} {\bibfnamefont {M.~A.}\ \bibnamefont
  {Schardt}}, \bibinfo {author} {\bibfnamefont {J.~S.}\ \bibnamefont {Frank}},
  \bibinfo {author} {\bibfnamefont {C.~M.}\ \bibnamefont {Hoffman}}, \bibinfo
  {author} {\bibfnamefont {R.~E.}\ \bibnamefont {Mischke}}, \bibinfo {author}
  {\bibfnamefont {D.~C.}\ \bibnamefont {Moir}}, \ and\ \bibinfo {author}
  {\bibfnamefont {P.~A.}\ \bibnamefont {Thompson}},\ }\href {\doibase
  10.1103/PhysRevD.23.639} {\bibfield  {journal} {\bibinfo  {journal} {Phys.
  Rev.}\ }\textbf {\bibinfo {volume} {D23}},\ \bibinfo {pages} {639} (\bibinfo
  {year} {1981})}\BibitemShut {NoStop}%
%%CITATION = PHRVA,D23,639;%%
\bibitem [{\citenamefont {Samios}(1961)}]{Samios:1961zz}%
  \BibitemOpen
  \bibfield  {author} {\bibinfo {author} {\bibfnamefont {N.~P.}\ \bibnamefont
  {Samios}},\ }\href {\doibase 10.1103/PhysRev.121.275} {\bibfield  {journal}
  {\bibinfo  {journal} {Phys. Rev.}\ }\textbf {\bibinfo {volume} {121}},\
  \bibinfo {pages} {275} (\bibinfo {year} {1961})}\BibitemShut {NoStop}%
%%CITATION = PHRVA,121,275;%%
\bibitem [{\citenamefont {Kampf}\ and\ \citenamefont
  {Moussallam}(2009)}]{Kampf:2009tk}%
  \BibitemOpen
  \bibfield  {author} {\bibinfo {author} {\bibfnamefont {K.}~\bibnamefont
  {Kampf}}\ and\ \bibinfo {author} {\bibfnamefont {B.}~\bibnamefont
  {Moussallam}},\ }\href {\doibase 10.1103/PhysRevD.79.076005} {\bibfield
  {journal} {\bibinfo  {journal} {Phys. Rev.}\ }\textbf {\bibinfo {volume}
  {D79}},\ \bibinfo {pages} {076005} (\bibinfo {year} {2009})},\ \Eprint
  {http://arxiv.org/abs/0901.4688} {arXiv:0901.4688 [hep-ph]} \BibitemShut
  {NoStop}%
%%CITATION = ARXIV:0901.4688;%%
\bibitem [{\citenamefont {Behrend}\ \emph {et~al.}(1991)\citenamefont {Behrend}
  \emph {et~al.}}]{Behrend:1990sr}%
  \BibitemOpen
  \bibfield  {author} {\bibinfo {author} {\bibfnamefont {H.~J.}\ \bibnamefont
  {Behrend}} \emph {et~al.} (\bibinfo {collaboration} {CELLO}),\ }\href
  {\doibase 10.1007/BF01549692} {\bibfield  {journal} {\bibinfo  {journal} {Z.
  Phys.}\ }\textbf {\bibinfo {volume} {C49}},\ \bibinfo {pages} {401} (\bibinfo
  {year} {1991})}\BibitemShut {NoStop}%
%%CITATION = ZEPYA,C49,401;%%
\bibitem [{\citenamefont {{Bergstr\"{o}m}}(1983)}]{Bergstrom:1982wk}%
  \BibitemOpen
  \bibfield  {author} {\bibinfo {author} {\bibfnamefont {L.}~\bibnamefont
  {{Bergstr\"{o}m}}},\ }\href {\doibase 10.1007/BF01573215} {\bibfield
  {journal} {\bibinfo  {journal} {Z. Phys.}\ }\textbf {\bibinfo {volume}
  {C20}},\ \bibinfo {pages} {135} (\bibinfo {year} {1983})}\BibitemShut
  {NoStop}%
%%CITATION = ZEPYA,C20,135;%%
\bibitem [{\citenamefont {Va\v{s}ko}\ and\ \citenamefont
  {Novotn\'y}(2011)}]{Vasko:2011pi}%
  \BibitemOpen
  \bibfield  {author} {\bibinfo {author} {\bibfnamefont {P.}~\bibnamefont
  {Va\v{s}ko}}\ and\ \bibinfo {author} {\bibfnamefont {J.}~\bibnamefont
  {Novotn\'y}},\ }\href {\doibase 10.1007/JHEP10(2011)122} {\bibfield
  {journal} {\bibinfo  {journal} {JHEP}\ }\textbf {\bibinfo {volume} {1110}},\
  \bibinfo {pages} {122} (\bibinfo {year} {2011})},\ \Eprint
  {http://arxiv.org/abs/1106.5956} {arXiv:1106.5956} \BibitemShut {NoStop}%
%%CITATION = ARXIV:1106.5956;%%
\bibitem [{\citenamefont {Husek}\ \emph {et~al.}(2014)\citenamefont {Husek},
  \citenamefont {Kampf},\ and\ \citenamefont {Novotn\'y}}]{Husek:2014tna}%
  \BibitemOpen
  \bibfield  {author} {\bibinfo {author} {\bibfnamefont {T.}~\bibnamefont
  {Husek}}, \bibinfo {author} {\bibfnamefont {K.}~\bibnamefont {Kampf}}, \ and\
  \bibinfo {author} {\bibfnamefont {J.}~\bibnamefont {Novotn\'y}},\ }\href
  {\doibase 10.1140/epjc/s10052-014-3010-4} {\bibfield  {journal} {\bibinfo
  {journal} {Eur. Phys. J.}\ }\textbf {\bibinfo {volume} {C74}},\ \bibinfo
  {pages} {3010} (\bibinfo {year} {2014})},\ \Eprint
  {http://arxiv.org/abs/1405.6927} {arXiv:1405.6927 [hep-ph]} \BibitemShut
  {NoStop}%
%%CITATION = ARXIV:1405.6927;%%
\bibitem [{\citenamefont {Husek}\ and\ \citenamefont
  {Leupold}(2015)}]{Husek:2015wta}%
  \BibitemOpen
  \bibfield  {author} {\bibinfo {author} {\bibfnamefont {T.}~\bibnamefont
  {Husek}}\ and\ \bibinfo {author} {\bibfnamefont {S.}~\bibnamefont
  {Leupold}},\ }\href {\doibase 10.1140/epjc/s10052-015-3778-x} {\bibfield
  {journal} {\bibinfo  {journal} {Eur. Phys. J.}\ }\textbf {\bibinfo {volume}
  {C75}},\ \bibinfo {pages} {586} (\bibinfo {year} {2015})},\ \Eprint
  {http://arxiv.org/abs/1507.00478} {arXiv:1507.00478 [hep-ph]} \BibitemShut
  {NoStop}%
%%CITATION = ARXIV:1507.00478;%%
\bibitem [{\citenamefont {Mikaelian}\ and\ \citenamefont
  {Smith}(1972)}]{Mikaelian:1972yg}%
  \BibitemOpen
  \bibfield  {author} {\bibinfo {author} {\bibfnamefont {K.~O.}\ \bibnamefont
  {Mikaelian}}\ and\ \bibinfo {author} {\bibfnamefont {J.}~\bibnamefont
  {Smith}},\ }\href {\doibase 10.1103/PhysRevD.5.1763} {\bibfield  {journal}
  {\bibinfo  {journal} {Phys. Rev.}\ }\textbf {\bibinfo {volume} {D5}},\
  \bibinfo {pages} {1763} (\bibinfo {year} {1972})}\BibitemShut {NoStop}%
%%CITATION = PHRVA,D5,1763;%%
\bibitem [{\citenamefont {Kampf}\ \emph {et~al.}(2006)\citenamefont {Kampf},
  \citenamefont {Knecht},\ and\ \citenamefont {Novotn\'y}}]{Kampf:2005tz}%
  \BibitemOpen
  \bibfield  {author} {\bibinfo {author} {\bibfnamefont {K.}~\bibnamefont
  {Kampf}}, \bibinfo {author} {\bibfnamefont {M.}~\bibnamefont {Knecht}}, \
  and\ \bibinfo {author} {\bibfnamefont {J.}~\bibnamefont {Novotn\'y}},\ }\href
  {\doibase 10.1140/epjc/s2005-02466-7} {\bibfield  {journal} {\bibinfo
  {journal} {Eur. Phys. J.}\ }\textbf {\bibinfo {volume} {C46}},\ \bibinfo
  {pages} {191} (\bibinfo {year} {2006})},\ \Eprint
  {http://arxiv.org/abs/hep-ph/0510021} {arXiv:hep-ph/0510021 [hep-ph]}
  \BibitemShut {NoStop}%
%%CITATION = HEP-PH/0510021;%%
\bibitem [{\citenamefont {Hoferichter}\ \emph
  {et~al.}(2018{\natexlab{a}})\citenamefont {Hoferichter}, \citenamefont
  {Hoid}, \citenamefont {Kubis}, \citenamefont {Leupold},\ and\ \citenamefont
  {Schneider}}]{Hoferichter:2018dmo}%
  \BibitemOpen
  \bibfield  {author} {\bibinfo {author} {\bibfnamefont {M.}~\bibnamefont
  {Hoferichter}}, \bibinfo {author} {\bibfnamefont {B.-L.}\ \bibnamefont
  {Hoid}}, \bibinfo {author} {\bibfnamefont {B.}~\bibnamefont {Kubis}},
  \bibinfo {author} {\bibfnamefont {S.}~\bibnamefont {Leupold}}, \ and\
  \bibinfo {author} {\bibfnamefont {S.~P.}\ \bibnamefont {Schneider}},\ }\href
  {\doibase 10.1103/PhysRevLett.121.112002} {\bibfield  {journal} {\bibinfo
  {journal} {Phys. Rev. Lett.}\ }\textbf {\bibinfo {volume} {121}},\ \bibinfo
  {pages} {112002} (\bibinfo {year} {2018}{\natexlab{a}})},\ \Eprint
  {http://arxiv.org/abs/1805.01471} {arXiv:1805.01471 [hep-ph]} \BibitemShut
  {NoStop}%
%%CITATION = ARXIV:1805.01471;%%
\bibitem [{\citenamefont {Hoferichter}\ \emph
  {et~al.}(2018{\natexlab{b}})\citenamefont {Hoferichter}, \citenamefont
  {Hoid}, \citenamefont {Kubis}, \citenamefont {Leupold},\ and\ \citenamefont
  {Schneider}}]{Hoferichter:2018kwz}%
  \BibitemOpen
  \bibfield  {author} {\bibinfo {author} {\bibfnamefont {M.}~\bibnamefont
  {Hoferichter}}, \bibinfo {author} {\bibfnamefont {B.-L.}\ \bibnamefont
  {Hoid}}, \bibinfo {author} {\bibfnamefont {B.}~\bibnamefont {Kubis}},
  \bibinfo {author} {\bibfnamefont {S.}~\bibnamefont {Leupold}}, \ and\
  \bibinfo {author} {\bibfnamefont {S.~P.}\ \bibnamefont {Schneider}},\ }\href
  {\doibase 10.1007/JHEP10(2018)141} {\bibfield  {journal} {\bibinfo  {journal}
  {JHEP}\ }\textbf {\bibinfo {volume} {10}},\ \bibinfo {pages} {141} (\bibinfo
  {year} {2018}{\natexlab{b}})},\ \Eprint {http://arxiv.org/abs/1808.04823}
  {arXiv:1808.04823 [hep-ph]} \BibitemShut {NoStop}%
%%CITATION = ARXIV:1808.04823;%%
\bibitem [{\citenamefont {Hoferichter}\ \emph {et~al.}(2014)\citenamefont
  {Hoferichter}, \citenamefont {Kubis}, \citenamefont {Leupold}, \citenamefont
  {Niecknig},\ and\ \citenamefont {Schneider}}]{Hoferichter:2014vra}%
  \BibitemOpen
  \bibfield  {author} {\bibinfo {author} {\bibfnamefont {M.}~\bibnamefont
  {Hoferichter}}, \bibinfo {author} {\bibfnamefont {B.}~\bibnamefont {Kubis}},
  \bibinfo {author} {\bibfnamefont {S.}~\bibnamefont {Leupold}}, \bibinfo
  {author} {\bibfnamefont {F.}~\bibnamefont {Niecknig}}, \ and\ \bibinfo
  {author} {\bibfnamefont {S.~P.}\ \bibnamefont {Schneider}},\ }\href {\doibase
  10.1140/epjc/s10052-014-3180-0} {\bibfield  {journal} {\bibinfo  {journal}
  {Eur. Phys. J.}\ }\textbf {\bibinfo {volume} {C74}},\ \bibinfo {pages} {3180}
  (\bibinfo {year} {2014})},\ \Eprint {http://arxiv.org/abs/1410.4691}
  {arXiv:1410.4691 [hep-ph]} \BibitemShut {NoStop}%
%%CITATION = ARXIV:1410.4691;%%
\bibitem [{\citenamefont {Masjuan}\ and\ \citenamefont
  {Sanchez-Puertas}(2017)}]{Masjuan:2017tvw}%
  \BibitemOpen
  \bibfield  {author} {\bibinfo {author} {\bibfnamefont {P.}~\bibnamefont
  {Masjuan}}\ and\ \bibinfo {author} {\bibfnamefont {P.}~\bibnamefont
  {Sanchez-Puertas}},\ }\href {\doibase 10.1103/PhysRevD.95.054026} {\bibfield
  {journal} {\bibinfo  {journal} {Phys. Rev.}\ }\textbf {\bibinfo {volume}
  {D95}},\ \bibinfo {pages} {054026} (\bibinfo {year} {2017})},\ \Eprint
  {http://arxiv.org/abs/1701.05829} {arXiv:1701.05829 [hep-ph]} \BibitemShut
  {NoStop}%
%%CITATION = ARXIV:1701.05829;%%
\bibitem [{\citenamefont {Masjuan}(2012)}]{Masjuan:2012wy}%
  \BibitemOpen
  \bibfield  {author} {\bibinfo {author} {\bibfnamefont {P.}~\bibnamefont
  {Masjuan}},\ }\href {\doibase 10.1103/PhysRevD.86.094021} {\bibfield
  {journal} {\bibinfo  {journal} {Phys. Rev.}\ }\textbf {\bibinfo {volume}
  {D86}},\ \bibinfo {pages} {094021} (\bibinfo {year} {2012})},\ \Eprint
  {http://arxiv.org/abs/1206.2549} {arXiv:1206.2549 [hep-ph]} \BibitemShut
  {NoStop}%
%%CITATION = ARXIV:1206.2549;%%
\bibitem [{\citenamefont {Adlarson}\ \emph {et~al.}(2017)\citenamefont
  {Adlarson} \emph {et~al.}}]{Adlarson:2016ykr}%
  \BibitemOpen
  \bibfield  {author} {\bibinfo {author} {\bibfnamefont {P.}~\bibnamefont
  {Adlarson}} \emph {et~al.} (\bibinfo {collaboration} {A2}),\ }\href {\doibase
  10.1103/PhysRevC.95.025202} {\bibfield  {journal} {\bibinfo  {journal} {Phys.
  Rev.}\ }\textbf {\bibinfo {volume} {C95}},\ \bibinfo {pages} {025202}
  (\bibinfo {year} {2017})},\ \Eprint {http://arxiv.org/abs/1611.04739}
  {arXiv:1611.04739 [hep-ex]} \BibitemShut {NoStop}%
%%CITATION = ARXIV:1611.04739;%%
\bibitem [{\citenamefont {Berman}\ and\ \citenamefont
  {Geffen}(1960)}]{Berman:1960zz}%
  \BibitemOpen
  \bibfield  {author} {\bibinfo {author} {\bibfnamefont {S.}~\bibnamefont
  {Berman}}\ and\ \bibinfo {author} {\bibfnamefont {D.}~\bibnamefont
  {Geffen}},\ }\href {\doibase 10.1007/BF02733176} {\bibfield  {journal}
  {\bibinfo  {journal} {Nuovo Cim.}\ }\textbf {\bibinfo {volume} {18}},\
  \bibinfo {pages} {1192} (\bibinfo {year} {1960})}\BibitemShut {NoStop}%
%%CITATION = NUCIA,18,1192;%%
\bibitem [{\citenamefont {Landsberg}(1985)}]{Landsberg:1986fd}%
  \BibitemOpen
  \bibfield  {author} {\bibinfo {author} {\bibfnamefont {L.}~\bibnamefont
  {Landsberg}},\ }\href {\doibase 10.1016/0370-1573(85)90129-2} {\bibfield
  {journal} {\bibinfo  {journal} {Phys. Rept.}\ }\textbf {\bibinfo {volume}
  {128}},\ \bibinfo {pages} {301} (\bibinfo {year} {1985})}\BibitemShut
  {NoStop}%
%%CITATION = PRPLC,128,301;%%
\bibitem [{\citenamefont {Sakurai}(1969)}]{sakuraiVMD}%
  \BibitemOpen
  \bibfield  {author} {\bibinfo {author} {\bibfnamefont {J.~J.}\ \bibnamefont
  {Sakurai}},\ }\href@noop {} {\emph {\bibinfo {title} {Currents and Mesons}}}\
  (\bibinfo  {publisher} {University of Chicago Press},\ \bibinfo {address}
  {Chicago},\ \bibinfo {year} {1969})\BibitemShut {NoStop}%
\bibitem [{\citenamefont {Peris}\ \emph {et~al.}(1998)\citenamefont {Peris},
  \citenamefont {Perrottet},\ and\ \citenamefont {de~Rafael}}]{Peris:1998nj}%
  \BibitemOpen
  \bibfield  {author} {\bibinfo {author} {\bibfnamefont {S.}~\bibnamefont
  {Peris}}, \bibinfo {author} {\bibfnamefont {M.}~\bibnamefont {Perrottet}}, \
  and\ \bibinfo {author} {\bibfnamefont {E.}~\bibnamefont {de~Rafael}},\ }\href
  {\doibase 10.1088/1126-6708/1998/05/011} {\bibfield  {journal} {\bibinfo
  {journal} {JHEP}\ }\textbf {\bibinfo {volume} {05}},\ \bibinfo {pages} {011}
  (\bibinfo {year} {1998})},\ \Eprint {http://arxiv.org/abs/hep-ph/9805442}
  {arXiv:hep-ph/9805442 [hep-ph]} \BibitemShut {NoStop}%
%%CITATION = HEP-PH/9805442;%%
\bibitem [{\citenamefont {Knecht}\ \emph {et~al.}(1999)\citenamefont {Knecht},
  \citenamefont {Peris}, \citenamefont {Perrottet},\ and\ \citenamefont
  {de~Rafael}}]{Knecht:1999gb}%
  \BibitemOpen
  \bibfield  {author} {\bibinfo {author} {\bibfnamefont {M.}~\bibnamefont
  {Knecht}}, \bibinfo {author} {\bibfnamefont {S.}~\bibnamefont {Peris}},
  \bibinfo {author} {\bibfnamefont {M.}~\bibnamefont {Perrottet}}, \ and\
  \bibinfo {author} {\bibfnamefont {E.}~\bibnamefont {de~Rafael}},\ }\href
  {\doibase 10.1103/PhysRevLett.83.5230} {\bibfield  {journal} {\bibinfo
  {journal} {Phys. Rev. Lett.}\ }\textbf {\bibinfo {volume} {83}},\ \bibinfo
  {pages} {5230} (\bibinfo {year} {1999})},\ \Eprint
  {http://arxiv.org/abs/hep-ph/9908283} {arXiv:hep-ph/9908283 [hep-ph]}
  \BibitemShut {NoStop}%
%%CITATION = HEP-PH/9908283;%%
\bibitem [{FFm()}]{FFmodels}%
  \BibitemOpen
  \href@noop {} {}\bibinfo {note} {{The simplest models do not, for instance,
  satisfy some of the major high-energy constraints: VMD violates
  operator-product-expansion (OPE)
  constraints~\cite{Knecht:2001xc,Husek:2015wta} and the LMD model cannot
  satisfy the Brodsky--Lepage scaling limit~\cite{Lepage:1979zb,Lepage:1980fj}.
  Although the THS model by construction satisfies the mentioned
  constraints~\cite{Husek:2015wta} and surpasses in this sense the VMD and LMD
  models, it inevitably suffers from shortcomings common to the whole family:
  for instance a presence of systematic uncertainties at low
  energies~\cite{Golterman:2006gv} (see also Ref.~\cite{Masjuan:2007ay}) or, at
  in the high-energy region,
  $\lim_{q^2\to\infty}q^2\mathcal{F}^\text{THS}(p^2,q^2)$ not being finite for
  all $p^2$, as a direct consequence of satisfying
  OPE~\cite{Stefan}.}}\BibitemShut {Stop}%
\bibitem [{\citenamefont {Husek}\ \emph {et~al.}(2018)\citenamefont {Husek},
  \citenamefont {Kampf}, \citenamefont {Leupold},\ and\ \citenamefont
  {Novotn\'y}}]{Husek:2017vmo}%
  \BibitemOpen
  \bibfield  {author} {\bibinfo {author} {\bibfnamefont {T.}~\bibnamefont
  {Husek}}, \bibinfo {author} {\bibfnamefont {K.}~\bibnamefont {Kampf}},
  \bibinfo {author} {\bibfnamefont {S.}~\bibnamefont {Leupold}}, \ and\
  \bibinfo {author} {\bibfnamefont {J.}~\bibnamefont {Novotn\'y}},\ }\href
  {\doibase 10.1103/PhysRevD.97.096013} {\bibfield  {journal} {\bibinfo
  {journal} {Phys. Rev.}\ }\textbf {\bibinfo {volume} {D97}},\ \bibinfo {pages}
  {096013} (\bibinfo {year} {2018})},\ \Eprint
  {http://arxiv.org/abs/1711.11001} {arXiv:1711.11001 [hep-ph]} \BibitemShut
  {NoStop}%
%%CITATION = ARXIV:1711.11001;%%
\bibitem [{\citenamefont {Tupper}\ \emph {et~al.}(1983)\citenamefont {Tupper},
  \citenamefont {Grose},\ and\ \citenamefont {Samuel}}]{Tupper:1983uw}%
  \BibitemOpen
  \bibfield  {author} {\bibinfo {author} {\bibfnamefont {G.~B.}\ \bibnamefont
  {Tupper}}, \bibinfo {author} {\bibfnamefont {T.~R.}\ \bibnamefont {Grose}}, \
  and\ \bibinfo {author} {\bibfnamefont {M.~A.}\ \bibnamefont {Samuel}},\
  }\href {\doibase 10.1103/PhysRevD.28.2905} {\bibfield  {journal} {\bibinfo
  {journal} {Phys. Rev.}\ }\textbf {\bibinfo {volume} {D28}},\ \bibinfo {pages}
  {2905} (\bibinfo {year} {1983})}\BibitemShut {NoStop}%
%%CITATION = PHRVA,D28,2905;%%
\bibitem [{\citenamefont {Lambin}\ and\ \citenamefont
  {Pestieau}(1985)}]{Lambin:1985sb}%
  \BibitemOpen
  \bibfield  {author} {\bibinfo {author} {\bibfnamefont {M.}~\bibnamefont
  {Lambin}}\ and\ \bibinfo {author} {\bibfnamefont {J.}~\bibnamefont
  {Pestieau}},\ }\href {\doibase 10.1103/PhysRevD.31.211} {\bibfield  {journal}
  {\bibinfo  {journal} {Phys. Rev.}\ }\textbf {\bibinfo {volume} {D31}},\
  \bibinfo {pages} {211} (\bibinfo {year} {1985})}\BibitemShut {NoStop}%
%%CITATION = PHRVA,D31,211;%%
\bibitem [{\citenamefont {Tupper}(1987)}]{Tupper:1986yk}%
  \BibitemOpen
  \bibfield  {author} {\bibinfo {author} {\bibfnamefont {G.}~\bibnamefont
  {Tupper}},\ }\href {\doibase 10.1103/PhysRevD.35.1726} {\bibfield  {journal}
  {\bibinfo  {journal} {Phys. Rev.}\ }\textbf {\bibinfo {volume} {D35}},\
  \bibinfo {pages} {1726} (\bibinfo {year} {1987})}\BibitemShut {NoStop}%
%%CITATION = PHRVA,D35,1726;%%
\bibitem [{det()}]{details}%
  \BibitemOpen
  \href@noop {} {}\bibinfo {note} {{The value corresponding to the correction
  $\delta_\text{conv}^\text{BS}$ includes not only the part with neglected
  slope $\Delta
  R_{\text{conv},a_\pi^\delta=0}^\text{BS}=-0.1582(1)\times10^{-3}$, but newly
  the form-factor correction was taken into account: $\Delta
  R_{\text{slope}}^\text{BS}=0.00062(13)\times10^{-3}$. This was calculated
  based on the final part of Section V of~\cite{Husek:2015sma} and is (due to
  the smallness of $a_\pi$) compatible with the approach described in Section V
  of~\cite{Husek:2017vmo}.}}\BibitemShut {Stop}%
\bibitem [{\citenamefont {Abouzaid}\ \emph {et~al.}(2007)\citenamefont
  {Abouzaid} \emph {et~al.}}]{Abouzaid:2006kk}%
  \BibitemOpen
  \bibfield  {author} {\bibinfo {author} {\bibfnamefont {E.}~\bibnamefont
  {Abouzaid}} \emph {et~al.} (\bibinfo {collaboration} {KTeV}),\ }\href
  {\doibase 10.1103/PhysRevD.75.012004} {\bibfield  {journal} {\bibinfo
  {journal} {Phys. Rev.}\ }\textbf {\bibinfo {volume} {D75}},\ \bibinfo {pages}
  {012004} (\bibinfo {year} {2007})},\ \Eprint
  {http://arxiv.org/abs/hep-ex/0610072} {arXiv:hep-ex/0610072 [hep-ex]}
  \BibitemShut {NoStop}%
%%CITATION = HEP-EX/0610072;%%
\bibitem [{\citenamefont {Niclasen}(2006)}]{Niclasen:2006tz}%
  \BibitemOpen
  \bibfield  {author} {\bibinfo {author} {\bibfnamefont {R.}~\bibnamefont
  {Niclasen}},\ }\emph {\bibinfo {title} {{Measuring the branching ratio of the
  rare decay $\pi^0\to e^+e^-$}}},\ \href {\doibase 10.2172/892361} {Ph.D.
  thesis},\ \bibinfo  {school} {Colorado U.} (\bibinfo {year}
  {2006})\BibitemShut {NoStop}%
%%CITATION = FERMILAB-THESIS-2006-12;%%
\bibitem [{\citenamefont {Kampf}\ \emph {et~al.}(2018)\citenamefont {Kampf},
  \citenamefont {Novotn\'y},\ and\ \citenamefont
  {Sanchez-Puertas}}]{Kampf:2018wau}%
  \BibitemOpen
  \bibfield  {author} {\bibinfo {author} {\bibfnamefont {K.}~\bibnamefont
  {Kampf}}, \bibinfo {author} {\bibfnamefont {J.}~\bibnamefont {Novotn\'y}}, \
  and\ \bibinfo {author} {\bibfnamefont {P.}~\bibnamefont {Sanchez-Puertas}},\
  }\href {\doibase 10.1103/PhysRevD.97.056010} {\bibfield  {journal} {\bibinfo
  {journal} {Phys. Rev.}\ }\textbf {\bibinfo {volume} {D97}},\ \bibinfo {pages}
  {056010} (\bibinfo {year} {2018})},\ \Eprint
  {http://arxiv.org/abs/1801.06067} {arXiv:1801.06067 [hep-ph]} \BibitemShut
  {NoStop}%
%%CITATION = ARXIV:1801.06067;%%
\bibitem [{\citenamefont {Amsler}\ \emph {et~al.}(2008)\citenamefont {Amsler}
  \emph {et~al.}}]{Amsler:2008zzb}%
  \BibitemOpen
  \bibfield  {author} {\bibinfo {author} {\bibfnamefont {C.}~\bibnamefont
  {Amsler}} \emph {et~al.} (\bibinfo {collaboration} {Particle Data Group}),\
  }\href {\doibase 10.1016/j.physletb.2008.07.018} {\bibfield  {journal}
  {\bibinfo  {journal} {Phys. Lett.}\ }\textbf {\bibinfo {volume} {B667}},\
  \bibinfo {pages} {1} (\bibinfo {year} {2008})}\BibitemShut {NoStop}%
%%CITATION = PHLTA,B667,1;%%
\bibitem [{\citenamefont {Joseph}(1960)}]{Joseph:1960zz}%
  \BibitemOpen
  \bibfield  {author} {\bibinfo {author} {\bibfnamefont {D.}~\bibnamefont
  {Joseph}},\ }\href {\doibase 10.1007/BF02860383} {\bibfield  {journal}
  {\bibinfo  {journal} {Nuovo Cim.}\ }\textbf {\bibinfo {volume} {16}},\
  \bibinfo {pages} {997} (\bibinfo {year} {1960})}\BibitemShut {NoStop}%
%%CITATION = NUCIA,16,997;%%
\bibitem [{\citenamefont {Lautrup}\ and\ \citenamefont
  {Smith}(1971)}]{Lautrup:1971ew}%
  \BibitemOpen
  \bibfield  {author} {\bibinfo {author} {\bibfnamefont {B.}~\bibnamefont
  {Lautrup}}\ and\ \bibinfo {author} {\bibfnamefont {J.}~\bibnamefont
  {Smith}},\ }\href {\doibase 10.1103/PhysRevD.3.1122} {\bibfield  {journal}
  {\bibinfo  {journal} {Phys. Rev.}\ }\textbf {\bibinfo {volume} {D3}},\
  \bibinfo {pages} {1122} (\bibinfo {year} {1971})}\BibitemShut {NoStop}%
%%CITATION = PHRVA,D3,1122;%%
\bibitem [{\citenamefont {Low}(1958)}]{Low:1958sn}%
  \BibitemOpen
  \bibfield  {author} {\bibinfo {author} {\bibfnamefont {F.~E.}\ \bibnamefont
  {Low}},\ }\href {\doibase 10.1103/PhysRev.110.974} {\bibfield  {journal}
  {\bibinfo  {journal} {Phys. Rev.}\ }\textbf {\bibinfo {volume} {110}},\
  \bibinfo {pages} {974} (\bibinfo {year} {1958})}\BibitemShut {NoStop}%
%%CITATION = PHRVA,110,974;%%
\bibitem [{\citenamefont {Adler}\ and\ \citenamefont
  {Dothan}(1966)}]{Adler:1966gc}%
  \BibitemOpen
  \bibfield  {author} {\bibinfo {author} {\bibfnamefont {S.~L.}\ \bibnamefont
  {Adler}}\ and\ \bibinfo {author} {\bibfnamefont {Y.}~\bibnamefont {Dothan}},\
  }\href {\doibase 10.1103/PhysRev.151.1267} {\bibfield  {journal} {\bibinfo
  {journal} {Phys. Rev.}\ }\textbf {\bibinfo {volume} {151}},\ \bibinfo {pages}
  {1267} (\bibinfo {year} {1966})}\BibitemShut {NoStop}%
%%CITATION = PHRVA,151,1267;%%
\bibitem [{\citenamefont {Pestleau}(1967)}]{Pestleau:1967snm}%
  \BibitemOpen
  \bibfield  {author} {\bibinfo {author} {\bibfnamefont {J.}~\bibnamefont
  {Pestleau}},\ }\href {\doibase 10.1103/PhysRev.160.1555} {\bibfield
  {journal} {\bibinfo  {journal} {Phys. Rev.}\ }\textbf {\bibinfo {volume}
  {160}},\ \bibinfo {pages} {1555} (\bibinfo {year} {1967})}\BibitemShut
  {NoStop}%
%%CITATION = PHRVA,160,1555;%%
\bibitem [{\citenamefont {Abouzaid}\ \emph {et~al.}(2008)\citenamefont
  {Abouzaid} \emph {et~al.}}]{Abouzaid:2008cd}%
  \BibitemOpen
  \bibfield  {author} {\bibinfo {author} {\bibfnamefont {E.}~\bibnamefont
  {Abouzaid}} \emph {et~al.} (\bibinfo {collaboration} {KTeV}),\ }\href
  {\doibase 10.1103/PhysRevLett.100.182001} {\bibfield  {journal} {\bibinfo
  {journal} {Phys. Rev. Lett.}\ }\textbf {\bibinfo {volume} {100}},\ \bibinfo
  {pages} {182001} (\bibinfo {year} {2008})},\ \Eprint
  {http://arxiv.org/abs/0802.2064} {arXiv:0802.2064 [hep-ex]} \BibitemShut
  {NoStop}%
%%CITATION = ARXIV:0802.2064;%%
\bibitem [{\citenamefont {Knecht}\ and\ \citenamefont
  {Nyffeler}(2001)}]{Knecht:2001xc}%
  \BibitemOpen
  \bibfield  {author} {\bibinfo {author} {\bibfnamefont {M.}~\bibnamefont
  {Knecht}}\ and\ \bibinfo {author} {\bibfnamefont {A.}~\bibnamefont
  {Nyffeler}},\ }\href {\doibase 10.1007/s100520100755} {\bibfield  {journal}
  {\bibinfo  {journal} {Eur. Phys. J.}\ }\textbf {\bibinfo {volume} {C21}},\
  \bibinfo {pages} {659} (\bibinfo {year} {2001})},\ \Eprint
  {http://arxiv.org/abs/hep-ph/0106034} {arXiv:hep-ph/0106034 [hep-ph]}
  \BibitemShut {NoStop}%
%%CITATION = HEP-PH/0106034;%%
\bibitem [{\citenamefont {Lepage}\ and\ \citenamefont
  {Brodsky}(1979)}]{Lepage:1979zb}%
  \BibitemOpen
  \bibfield  {author} {\bibinfo {author} {\bibfnamefont {G.~P.}\ \bibnamefont
  {Lepage}}\ and\ \bibinfo {author} {\bibfnamefont {S.~J.}\ \bibnamefont
  {Brodsky}},\ }\href {\doibase 10.1016/0370-2693(79)90554-9} {\bibfield
  {journal} {\bibinfo  {journal} {Phys. Lett.}\ }\textbf {\bibinfo {volume}
  {87B}},\ \bibinfo {pages} {359} (\bibinfo {year} {1979})}\BibitemShut
  {NoStop}%
%%CITATION = PHLTA,87B,359;%%
\bibitem [{\citenamefont {Lepage}\ and\ \citenamefont
  {Brodsky}(1980)}]{Lepage:1980fj}%
  \BibitemOpen
  \bibfield  {author} {\bibinfo {author} {\bibfnamefont {G.~P.}\ \bibnamefont
  {Lepage}}\ and\ \bibinfo {author} {\bibfnamefont {S.~J.}\ \bibnamefont
  {Brodsky}},\ }\href {\doibase 10.1103/PhysRevD.22.2157} {\bibfield  {journal}
  {\bibinfo  {journal} {Phys. Rev.}\ }\textbf {\bibinfo {volume} {D22}},\
  \bibinfo {pages} {2157} (\bibinfo {year} {1980})}\BibitemShut {NoStop}%
%%CITATION = PHRVA,D22,2157;%%
\bibitem [{\citenamefont {Golterman}\ and\ \citenamefont
  {Peris}(2006)}]{Golterman:2006gv}%
  \BibitemOpen
  \bibfield  {author} {\bibinfo {author} {\bibfnamefont {M.}~\bibnamefont
  {Golterman}}\ and\ \bibinfo {author} {\bibfnamefont {S.}~\bibnamefont
  {Peris}},\ }\href {\doibase 10.1103/PhysRevD.74.096002} {\bibfield  {journal}
  {\bibinfo  {journal} {Phys. Rev.}\ }\textbf {\bibinfo {volume} {D74}},\
  \bibinfo {pages} {096002} (\bibinfo {year} {2006})},\ \Eprint
  {http://arxiv.org/abs/hep-ph/0607152} {arXiv:hep-ph/0607152 [hep-ph]}
  \BibitemShut {NoStop}%
%%CITATION = HEP-PH/0607152;%%
\bibitem [{\citenamefont {Masjuan}\ and\ \citenamefont
  {Peris}(2007)}]{Masjuan:2007ay}%
  \BibitemOpen
  \bibfield  {author} {\bibinfo {author} {\bibfnamefont {P.}~\bibnamefont
  {Masjuan}}\ and\ \bibinfo {author} {\bibfnamefont {S.}~\bibnamefont
  {Peris}},\ }\href {\doibase 10.1088/1126-6708/2007/05/040} {\bibfield
  {journal} {\bibinfo  {journal} {JHEP}\ }\textbf {\bibinfo {volume} {05}},\
  \bibinfo {pages} {040} (\bibinfo {year} {2007})},\ \Eprint
  {http://arxiv.org/abs/0704.1247} {arXiv:0704.1247 [hep-ph]} \BibitemShut
  {NoStop}%
%%CITATION = ARXIV:0704.1247;%%
\bibitem [{\citenamefont {Leupold}\ and\ \citenamefont {Kubis}()}]{Stefan}%
  \BibitemOpen
  \bibfield  {author} {\bibinfo {author} {\bibfnamefont {S.}~\bibnamefont
  {Leupold}}\ and\ \bibinfo {author} {\bibfnamefont {B.}~\bibnamefont
  {Kubis}},\ }\href@noop {} {}\bibinfo {howpublished} {personal
  communication}\BibitemShut {NoStop}%
\end{thebibliography}
%\input{Dalitz.bbl}

%

\end{document}